\documentclass[twocolumn,preprintnumbers,amsmath,amssymb]{revtex4}
\usepackage{graphicx}% Include figure files
\usepackage{dcolumn}% Align table columns on decimal point
\usepackage{bm}% bold math
\usepackage{color}
\usepackage{setspace}
\begin{document}

%\preprint{AIP/*****}

\title{Signature of magnon Nernst effect in an antiferromagnetic insulator}% Force line breaks with \\

\author{Y. Shiomi$^{\, 1}$}
%\email{shiomi@imr.tohoku.ac.jp}
\author{R. Takashima$^{\, 2}$}
\author{E. Saitoh$^{\, 1,3,4,5}$}

%\affiliation{$^{1}$
%Department of Physics, Graduate School of Science, Tohoku University, Sendai, 980-8578, Japan}
\affiliation{$^{1}$
Institute for Materials Research, Tohoku University, Sendai 980-8577, Japan }
\affiliation{$^{2}$
Department of Physics, Kyoto University, Kyoto 606-8502, Japan}
\affiliation{$^{3}$
WPI Advanced Institute for Materials Research, Tohoku University, Sendai 980-8577, Japan
}
\affiliation{$^{4}$
Advanced Science Research Center, Japan Atomic Energy Agency, Tokai 319-1195, Japan
}
\affiliation{$^{5}$
Center for Spintronics Research Network, Tohoku University, Sendai 980-8577, Japan
}

\date{\today}% It is always \today, today,
             %  but any date may be explicitly specified

\begin{abstract}
A magnon Nernst effect, an antiferromagnetic analogue of the magnon Hall effect in ferromagnetic insulators, has been studied experimentally for a layered antiferromagnetic insulator MnPS$_{3}$ in contact with two Pt strips. Thermoelectric voltage in the Pt strips grown on MnPS$_{3}$ single crystals exhibits non-monotonic temperature dependence at low temperatures, which cannot be explained by electronic origins in Pt but can be ascribed to the inverse spin Hall voltage induced by a magnon Nernst effect. Control of antiferromagnetic domains in the MnPS$_{3}$ crystal by magnetoelectric cooling is found to modulate the low-temperature thermoelectric voltage in Pt, which corroborates the emergence of the magnon Nernst effect in Pt$\mid$MnPS$_{3}$ hybrid structures.       
\end{abstract}

%\pacs{72.25.-b, 75.76.+j, 74.25.Fy, 74.25.Qt, 74.72.Bk, 72.20.Pa}% PACS, the Physics and Astronomy
                             % Classification Scheme.
%\keywords{Suggested keywords}%Use showkeys class option if keyword
                              %display desired
\maketitle

%\section*{Introduction}
Berry phase is a fundamental concept in solid state physics and responsible for a spectrum of physical phenomena \cite{berry-phase-review}. One pronounced example caused by Berry phase is Hall effects of electrons. The Berry curvature of electrons is made manifest as transverse velocity of the electrons, which gives rise to various Hall effects, {\it e.g.} anomalous Hall effects, topological Hall effects, and spin Hall effects \cite{berry-phase-review, AHE-review, SHE-review}. In semi-classical theory, the Berry curvature can be regarded as an effective magnetic flux for conduction electrons, and the related effective Lorentz force bends the trajectory of moving electrons to the Hall direction \cite{berry-phase-review, AHE-review, SHE-review}. \par  

\begin{figure}[t]
\begin{center}
\includegraphics[width=7.5cm]{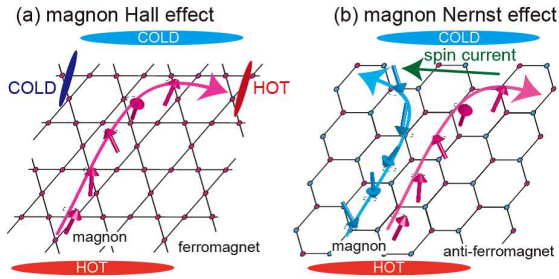}
\caption{Schematics of (a) the magnon Hall effect in kagome ferromagnets and (b) the magnon Nernst effect in honeycomb antiferromagnets.    } 
\label{fig1}
\end{center}
\end{figure}

Recently, the Berry phase concept has been expanded to magnon transport in ferromagnetic insulators \cite{katsura, matsumoto}. In insulating magnets, the spin transport is governed by low-energy spin excitations, {\it i.e.} magnons. The ring exchange process on ferromagnetic lattices leads to a Berry phase effect in magnon transport \cite{katsura}. With certain types of lattice geometry and magnetic order, the net fictitious magnetic field due to the magnon Berry curvature survives, and there occurs the magnon Hall effect \cite{katsura, matsumoto}. As shown in Fig. \ref{fig1}(a), the magnon Hall current produces temperature gradient along the Hall direction, leading to finite thermal Hall conductivity. The magnon Hall effect has been demonstrated experimentally by the measurement of the thermal Hall effect (Righi-Leduc effect) for a pyrochlore ferromagnet Lu$_{2}$V$_{2}$O$_{7}$ \cite{onose} and other magnetic insulators \cite{ideue, ong-mhe, ideue-nmat}. 
\par

In this letter, the topological magnon transport induced by the Berry curvature is experimentally expanded to antiferromagnetic insulators: magnon-mediated spin Nernst effect, dubbed as magnon Nernst effect \cite{okamoto, kovalev}. In a honeycomb antiferromagnet in the presence of the Dzyaloshinskii-Moriya (DM) interaction, a longitudinal temperature gradient can give rise to spin currents along the Hall direction, realizing a magnon Nernst effect \cite{okamoto, kovalev}. This effect can be viewed as an antiferromagnetic analogue of the magnon Hall effect in ferromagnetic insulators, which stems from a fictitious magnetic flux due to the magnon Berry curvature. As shown in Fig. \ref{fig1}(b), in a honeycomb antiferromagnet \cite{okamoto, kovalev}, both up-spin and down-spin magnons exhibit the magnon Hall effect under an applied inplane temperature gradient, but magnons with opposite spins flow in opposite transverse directions. The thermal Hall current thereby cancels out, while a nonzero spin current emerges in the Hall direction. Generation of spin currents devoid of a thermal current without external magnetic fields should be a significant merit of antiferromagnetic spintronics \cite{AFspintronics}.
\par

As a material candidate to realize magnon Nernst effects, monolayer MnPS$_{3}$ has been theoretically discussed \cite{okamoto} and non-monotonic temperature dependence of the magnon Nernst coefficient was predicted based on a low-temperature spin-wave approximation [see Fig. \ref{fig4}(d)]. Though growth of the monolayer MnPS$_{3}$ is experimentally challenging, the similar effect is expected in bulk MnPS$_{3}$ because of the two-dimensional crystal structure with tiny interlayer magnetic couplings \cite{wildes}. Bulk single crystals of MnPS$_{3}$ have been intensively studied because of the two-dimensional character in the magnetic properties \cite{kurosawa} and the easy intercalation of lithium and molecules \cite{brec}. In order to study magnon Nernst effects expected in MnPS$_{3}$, we here prepare bulk single crystals of MnPS$_{3}$ and employ the technique of the inverse spin Hall effect \cite{saitoh} for electrical detection of the transverse spin current induced by the magnon Nernst effect. For Pt films grown on the edges of MnPS$_{3}$ single crystals, we have observed non-monotonic temperature dependence of the thermoelectric voltage and also its dependence on antiferromagnetic domains of MnPS$_{3}$. These results indicate that the magnon Nernst effect induces inverse spin Hall voltage in the Pt$\mid$MnPS$_{3}$ structure.          
\par

\begin{figure}[t]
\begin{center}
\includegraphics[width=7.5cm]{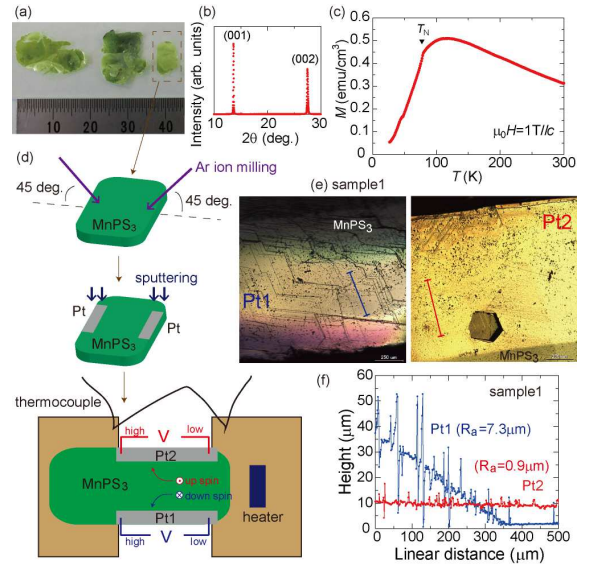}
\caption{(a) Examples of MnPS$_{3}$ single crystals. (b) X-ray diffraction pattern of a MnPS$_{3}$ single crystal. (c) Temperature ($T$) dependence of the magnetization ($M$) for a MnPS$_{3}$ single crystal. The measurement was performed under magnetic fields of $1$ T applied along the $c$ axis. (d) Flow chart of sample setup in the magnon Nernst effect measurement. After the surface treatment by Ar-ion milling processes, $5$-nm-thick Pt films were deposited on the MnPS$_{3}$ surfaces. The Pt$\mid$MnPS$_{3}$ samples were fixed on two heat sinks for the measurement of thermoelectric voltages. (e) Surface images taken with a laser microscope for Pt1 and Pt2 of Sample1. (f) Distance distribution of surface heights for Pt1 and Pt2 of Sample1. This graph is obtained from the images shown in (e).} 
\label{fig2}
\end{center}
\end{figure}

Bulk single crystals of MnPS$_{3}$ were grown by a chemical vapor transport method following a previous report \cite{date}. Stoichiometric amounts totaling $5$ g of Mn, P, and S elements were sealed into an evacuated quartz tube. The tube was placed in a horizontal three-zone furnace and heated slowly up to $630$-$680$ $^{\circ}$C; the temperature of the charge region was set at $680$ $^{\circ}$C and that of the growth region at $630$ $^{\circ}$C. A number of plate-like single crystals of MnPS$_{3}$ were obtained in the growth region in 100 hours. As shown in Fig. \ref{fig2}(a), the MnPS$_{3}$ single crystals are optically transparent with a green color. The largest surfaces of the crystals were determined by x-ray diffraction measurements to be the $ab$ plane, as shown in Fig. \ref{fig2}(b). Manganese ions (Mn$^{2+}$) form a honeycomb lattice within the $ab$ plane and the honeycomb lattices are stacked along the $c$ direction with a weak Van-der-Waals interlayer coupling \cite{brec}.   
\par

The MnPS$_{3}$ single crystals are highly insulating below room temperature, which is suitable for studying spin-current generation free from magneto-transport effects inside MnPS$_{3}$. The magnetic property of MnPS$_{3}$ was studied using a SQUID magnetometer (Magnetic Property Measurement System; Quantum Design, Inc.). The temperature ($T$) dependence of the magnetization, $M$, for a MnPS$_{3}$ single crystal is shown in Fig. \ref{fig2}(c). Here, the external magnetic field ($H$) of $1$ T was applied perpendicular to the crystal plane. With decreasing $T$ from $300$ K, $M$ increases and shows a broad peak around $110$ K, reflecting a short range order of Mn$^{2+}$ spins \cite{date}. The magnetization sharply decreases below $80$ K, which corresponds to the antiferromagnetic ordering temperature of Mn$^{2+}$ spins ($T_{N}$). Each Mn$^{2+}$ spin in the $ab$ plane is coupled antiferromagnetically with the nearest neighbors and coupled ferromagnetically with the interlayer neighbors \cite{date}. The observed $M$-$T$ curve [Fig. \ref{fig2}(c)] is similar to that reported previously \cite{date}, which supports the high quality of our single crystals.       
\par

Seemingly homogeneous crystals of MnPS$_{3}$ were selected and used for the measurement of the magnon Nernst effect. Since MnPS$_{3}$ single crystals are soft and fragile, the surface treatment by mechanical polish is difficult, and then Ar-ion milling was applied. As illustrated in Fig. \ref{fig2}(d), MnPS$_{3}$ crystals were irradiated with Ar ions at an acceleration voltage of $500$ V at $45^{\circ}$ angles to the crystal planes for $30$ minutes on a water cooled sample holder. The irradiation was done in the intervals of $10$ minutes with a pause of time longer than $10$ minutes in order to avoid sample damages. On both the edges of MnPS$_{3}$ surfaces, $5$-nm-thick Pt films (with the size of $\sim 4 \times 1$ mm$^{2}$) were sputtered at room temperature in an Ar atmosphere, as shown in Fig. \ref{fig2}(d). 
\par

The obtained Pt$\mid$MnPS$_{3}$ structures were fixed on two heat sinks using GE varnish, as shown in Fig. \ref{fig2}(d); on one heat sink, a $1$k${\rm \Omega}$ chip resistor was put to generate temperature gradient in an inplane direction. The temperature difference between the two heat sinks, $\Delta T$, was measured with a couple of type-E thermocouples. The thermoelectric voltage, $V$, for the two Pt strips (denoted by Pt1 and Pt2 hereafter) was recorded with Keithly 2182A nanovoltmeters. The measurements were performed in a cryogenic probe station in the temperature range between $10$ K and $200$ K without external magnetic fields. Note that the values of $\Delta T$ in the thermoelectric measurements are shown in Supplemental Material (SM) \cite{SM}.  
\par

\begin{figure}[t]
\begin{center}
\includegraphics[width=7.5cm]{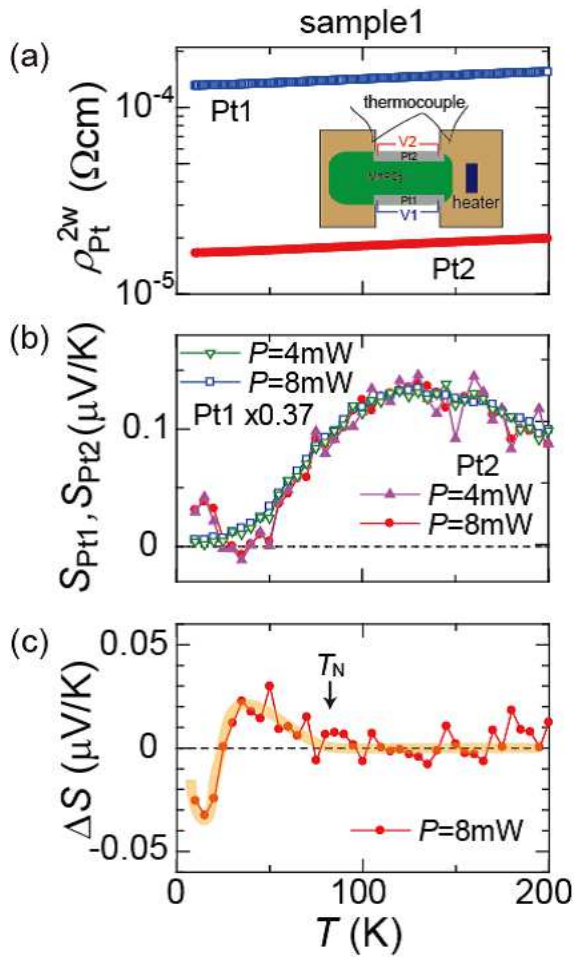}
\caption{Temperature ($T$) dependence of (a) two-wire resistivity ($\rho_{\rm Pt}^{\rm 2w}$) for Pt1 and Pt2 of Sample1, (b) $V/\Delta T$ for Pt1 and Pt2 of Sample1 ($S_{\rm Pt1}$ and $S_{\rm Pt2}$, respectively), and (c) the anomalous thermoelectric contribution defined as $\Delta S \equiv 0.37 \times S_{\rm Pt1} - S_{\rm Pt2}$ for Sample1. In (b), $0.37 \times S_{\rm Pt1}$ at each $T$ is plotted for Pt1. Here, $P$ in (b) and (c) denotes the power level applied to the heater.  } 
\label{fig3}
\end{center}
\end{figure}

Figure \ref{fig3} shows a set of experimental results for Sample$1$ of Pt$\mid$MnPS$_{3}$. In Fig. \ref{fig3}(a), two-wire resistivity, $\rho_{\rm Pt}^{\rm 2w}$, measured between the voltage electrodes for Pt1 and Pt2 was presented. For Pt2, $\rho_{\rm Pt}^{\rm 2w}$ is lower than $2 \times 10^{-5}$ ${\rm \Omega cm}$, which is almost the same as that of Pt films grown on an oxidized-Si substrate in our sputtering condition \cite{otani}. By contrast, $\rho_{\rm Pt}^{\rm 2w}$ for Pt1 is over $1 \times 10^{-4}$ ${\rm \Omega cm}$, about ten times greater than that for Pt2. The difference in $\rho_{\rm Pt}^{\rm 2w}$ between Pt1 and Pt2 should be related with the roughness of the MnPS$_{3}$ surface, as shown in Figs. \ref{fig2}(e) and \ref{fig2}(f). Arithmetic average roughness ($R_{a}$) is $0.9$ ${\rm \mu}$m for Pt2$\mid$MnPS$_{3}$, while as large as $7.3$ ${\rm \mu}$m for Pt1$\mid$MnPS$_{3}$ in Fig. \ref{fig2}(e). The Pt2$\mid$MnPS$_{3}$ interface is much smoother than Pt1$\mid$MnPS$_{3}$; higher spin mixing conductance is expected at the Pt2$\mid$MnPS$_{3}$ interface.  
\par

Figure \ref{fig3}(b) shows $T$ dependence of the thermoelectric coefficient for Pt1 and Pt2 strips of Sample1, $S_{\rm Pt1}$ and $S_{\rm Pt2}$. Here, $S_{\rm Pt1}$ and $S_{\rm Pt2}$ are defined by the thermoelectric voltage $V$ divided by the temperature difference $\Delta T$ for Pt1 and Pt2 strips, respectively. Not only the Seebeck voltage of the Pt films but also the inverse spin Hall voltage induced by the magnon Nernst effect can contribute to $S_{\rm Pt1}$ and $S_{\rm Pt2}$ if it exists; note that the same sign of the magnon Nernst voltage is expected for Pt1 and Pt2 because the direction of spin polarization and also the direction of its flow are opposite between Pt1 and Pt2 [see Fig. \ref{fig2}(d)]. As shown in Fig. \ref{fig3}(b), $S_{\rm Pt1}$ and $S_{\rm Pt2}$ show a similar $T$ dependence at high temperatures above $T_{N}$, while the magnitude of $S_{\rm Pt1}$ is larger than that of $S_{\rm Pt2}$ (see also Fig. S1 \cite{SM}). $S_{\rm Pt1}$ and $S_{\rm Pt2}$ exhibit the positive sign and show a broad peak around $120$ K, which is consistent with the $T$ dependence of the Seebeck coefficient for Pt bulk samples \cite{Pt-seebeck}. Hence, at high temperatures above $T_{N}$, the conventional Seebeck effect well explains $S_{\rm Pt1}$ and $S_{\rm Pt2}$. 
\par

At low temperatures below $T_{N}$, however, $S_{\rm Pt1}$ and $S_{\rm Pt2}$ totally show different $T$ dependences. For high-resistive Pt1, $S_{\rm Pt1}$ monotonically decreases to zero as $T$ decreases, as expected by the Seebeck effect in Pt \cite{Pt-seebeck}. In contrast, $S_{\rm Pt2}$ for low-resistive Pt2 shows a sign change from positive to negative at about $30$ K, and then the sign changes again from negative to positive to exhibit a positive peak at $15$ K. This highly non-monotinic $T$ dependence of $S_{\rm Pt2}$ cannot be explained by electronic origins of Pt, such as a change in dominant carriers at low temperatures or the phonon drag; such a complex electronic structure which can explain the serial sign changes ({\it i.e.} positive-negative-positive sign changes) is not observed in Pt. Also, the Seebeck anomaly due to the phonon drag is observed as a single peak \cite{barnard}. Hence, the anomalous $T$ dependence observed in $S_{\rm pt2}$ below $T_{N}$ is not explained by the simple modulation of the Seebeck voltages of Pt at low temperatures.             
\par 

Notable is that the anomalous $T$ dependence of $S_{\rm Pt2}$ at low temperatures below $T_{N}$ is consistent with the theoretical prediction on the magnon Nernst effect for monolayer MnPS$_{3}$ \cite{okamoto}. In Fig. \ref{fig3}(c), the anomalous thermoelectric contribution in $S_{\rm Pt2}$ is separated from the conventional Seebeck effect by plotting $\Delta S \equiv 0.37 \times S_{\rm Pt1} - S_{\rm Pt2}$. As $T$ decreases below $T_{N}$, $\Delta S$ evolves and shows a broad positive peak around $30$ K, and then sharply drops to exhibit a sign change with a negative peak at $15$ K. This $T$ dependence of $\Delta S$ is consistent with that calculated theoretically for MnPS$_{3}$ in Fig. \ref{fig4}(d), where the sign change of the magnon Nernst voltage is predicted at $20$ K-$40$ K by the sign flip of the magnon Berry curvature across the von Hove singularities \cite{okamoto}. The sign-change temperature of about $25$ K corresponds to the DM parameter ($D$) of $\approx 0.3$ meV \cite{SM}.    
\par

\begin{figure}[t]
\begin{center}
\includegraphics[width=7.5cm]{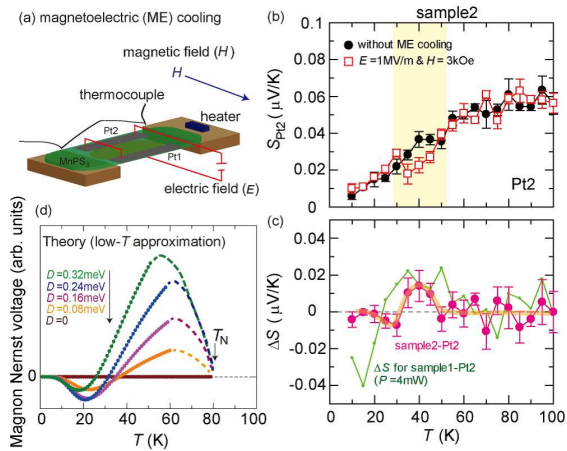}
\caption{(a) A schematic illustration of experimental configuration of magnetoelectric (ME) cooling. The electric field ($E$) is applied along the perpendicular-to-plane direction and the magnetic field ($H$) applied along an in-plane direction. (b) Temperature ($T$) dependence of $V/\Delta T$ for Pt2 ($S_{\rm Pt2}$) of Sample2 without ME cooling (circles) and after a ME cooling of $E= 1$ MV/m and $H= 3$ kOe (squares). (c) Temperature ($T$) dependence of the change of $S_{\rm Pt2}$ values by the ME cooling ($\Delta S$) for Pt2 of Sample2 (circles). The data of $\Delta S$ for Pt2 of Sample1 at $P=4$ mW is also shown for comparison. (d) Temperature ($T$) dependence of the magnon Nernst signal predicted theoretically for monolayer MnPS$_{3}$ [\onlinecite{okamoto}]. The calculation is only valid at low temperatures, and the magnon Nernst signal should vanish above $T_{N}$ of MnPS$_{3}$ as indicated by dotted curves. } 
\label{fig4}
\end{center}
\end{figure}

As for the indiscernible magnon Nernst voltage in the Pt1 strip of Sample1 [Fig. \ref{fig3}(b) and Fig. S1 \cite{SM}], several possibilities can be raised. First, sizable roughness of MnPS$_{3}$ crystals as seen in Figs. \ref{fig2}(e) and \ref{fig2}(f) should decrease the spin mixing conductance between Pt and MnPS$_{3}$ layers, resulting in the suppression of the magnon Nernst voltage. In our experiments, the magnitude of $\rho_{\rm Pt}^{\rm 2w}$ strongly depends on samples. It is notable that $\rho_{\rm Pt}^{\rm 2w}$ for Pt2 of Sample1 is the lowest among our Pt$\mid$MnPS$_{3}$ samples, which indicates high spin mixing conductance. Second, the magnon Nernst voltage is expected to be smaller than the Seebeck voltage of Pt. Large Seebeck coefficients as observed in Pt1 of Sample1 are not suitable for the detection of small magnon Nernst voltage at low temperatures. Third, mixed antiferromagnetic domains whose boundaries disturb magnon transports can make the detection of the magnon Nernst voltage difficult. 
\par

To reveal the effect of antiferromagnetic domains in MnPS$_{3}$ on the magnon Nernst effect, we have investigated the dependence of the thermoelectric voltage on magnetoelectric (ME) cooling for Sample2 in Figs. \ref{fig4}(a)-\ref{fig4}(c); here, $\rho_{\rm Pt}^{\rm 2w}$ and the Seebeck coefficient for Pt strips of Sample2 were relatively small among our samples (Figs. \ref{fig4} and S2 \cite{SM}), but the magnon Nernst voltage was not observed within our experimental accuracy. It is known that MnPS$_{3}$ exhibits a linear ME effect and that antiferromagnetic domains can be manipulated by cooling the sample under crossed magnetic and electric fields \cite{ferrotroidic}. Although different antiferromagnetic domains are expected to produce the magnon Nernst voltage of the same sign \cite{SM}, the multidomain states should reduce the magnon Nernst voltage because of disturbance of magnon transport by domain boundaries. Hence, the modulation of the thermoelectric voltage by ME cooling can corroborate the emergence of the magnon Nernst effect. As shown in Fig. \ref{fig4}(a), the MnPS$_{3}$ single crystal was submitted to electric and magnetic fields; the electric field ($E$) was applied along the perpendicular direction to the $ab$ plane, while the magnetic field ($H$) along an inplane direction perpendicular to the Pt strips. Here, the gate-voltage electrode on the bottom surface was formed using a conductive silver paste \cite{seki} and the resistance between Pt strips and the bottom electrode was over the measurable range of multimeters. After ME cooling from $200$ K down to $10$ K under the simultaneous action of $E= 1$ MV/m (corresponding to $40$ V) and $H= 3$ kOe, measurement of the thermoelectric voltages for Pt strips of Sample2 was performed without any fields in increasing-$T$ scans, as shown in Fig. \ref{fig4}(b). Obviously, the value of $S_{\rm Pt2}$ after the ME cooling is found to be less than that measured without the ME cooling in the $T$ range between $30$ and $50$ K, while it is less sensitive to the ME cooling in other $T$ ranges. The change in the thermoelectric voltage of Pt by the ME cooling indicates that the voltage related with the antiferromagnetic domains in MnPS$_{3}$ is included in the thermoelectric voltage of Pt.      
\par

The change of the $S_{\rm Pt2}$ values by the ME cooling is calculated for Pt2 of Sample2, and is plotted against the measurement temperature, as shown in Fig. \ref{fig4}(c). A dome-shaped positive thermoelectric signal is tangible between $30$ K and $50$ K. Notably, the magnitude and $T$-dependence of the $\Delta S$ peak for Sample2 are similar to those of $\Delta S$ observed in Sample1 [Figs. \ref{fig3}(c) and \ref{fig4}(c)]. Note that, in the measurement for Sample2, the magnitudes of $\Delta T$ were set to be $\sim 10$ K  at $10$ K (Fig. S2 \cite{SM}) to improve the signal to noise ratio. Hence, the low-$T$ negative peak as observed at $15$ K for Pt2 of Sample1 in Fig. \ref{fig3}(c) is smeared for Pt2 of Sample2 owing to heating effects. The modulation of the thermoelectric voltage by ME cooling corroborates the emergence of the magnon Nernst effect for Pt$\mid$MnPS$_{3}$. The magnitude of the inverse spin Hall voltage induced by the magnon Nernst effect is at most several tens nanovolts in the present experiments.         
\par

In summary, we measured thermoelectric voltages of Pt strips grown on the edges of MnPS$_{3}$ single crystals. The anomalous temperature dependence of the thermoelectric voltage for a low-resistive Pt strip on MnPS$_{3}$ is notably consistent with that of the magnon Nernst coefficient predicted in a theoretical work \cite{okamoto}. Furthermore, the modulation of the low-temperature thermoelectric voltage by magnetoelectric cooling was demonstrated, which shows that the thermoelectric voltage includes spin-related voltage signals dependent on antiferromagnetic domains of MnPS$_{3}$. The results signal the emergence of the magnon Nernst effect in the Pt$\mid$MnPS$_{3}$ samples. We hope that the present work will stimulate follow-up studies on the magnon Nernst effect in antiferromagnetic insulators to harmonize the spin caloritronics, antiferromagnetic spintronics, and topological spintronics.    
\par

%acknowledgement
We are grateful to S. Daimon for informing Y. S. of a theoretical paper on the magnon spin Nernst effect \cite{okamoto}. This work was supported by JST ERATOgSpin Quantum Rectification Projecth (JPMJER1402), JSPS KAKENHI (No. 17H04806, No. JP16H00977, No. 16K13827, and No. 15J01700), and MEXT (Innovative Area ``Nano Spin Conversion Science" (No. 26103005)).
\par

%
%\clearpage

%\clearpage

%\clearpage

%\clearpage


\begin{thebibliography}{99}
\bibitem{berry-phase-review}D. Xiao, M.-C. Chang, and Q. Niu, Rev. Mod. Phys. {\bf 82}, 1959 (2010).
\bibitem{AHE-review}N. Nagaosa, J. Sinova, S. Onoda, A. H. MacDonald, and N. P. Ong, Rev. Mod. Phys. {\bf 82}, 1539 (2010).
\bibitem{SHE-review}J. Sinova, S. O. Valenzuela, J. Wunderlich, C. H. Back, and T. Jungwirth, Rev. Mod. Phys. {\bf 87}, 1213 (2015).
\bibitem{katsura}H. Katsura, N. Nagaosa, and P. A. Lee, Phys. Rev. Lett. {\bf 104}, 066403 (2010). 
\bibitem{matsumoto}R. Matsumoto and S. Murakami, Phys. Rev. Lett. {\bf 106}, 197202 (2011). 
\bibitem{onose}Y. Onose, {\it et al.} Science {\bf 329}, 297 (2010). 
\bibitem{ideue}T. Ideue, {\it et al.} Phys. Rev. B {\bf 85}, 134411 (2012).
\bibitem{ong-mhe}M. Hirschberger, R. Chisnell, Y. S. Lee, and N. P. Ong, Phys. Rev. Lett. {\bf 115}, 106603 (2015).
\bibitem{ideue-nmat}T. Ideue, T. Kurumaji, S. Ishiwata, and Y. Tokura, Nature. Mater. {\it in press. (advance online publication)} (2017).
\bibitem{okamoto}R. Cheng, S. Okamoto, and D. Xiao, Phys. Rev. Lett. {\bf 117}, 217202 (2016).
\bibitem{kovalev}V. A. Zyuzin and A. A. Kovalev, Phys. Rev. Lett. {\bf 117}, 217203 (2016).
\bibitem{AFspintronics}O. Gomonay, T. Jungwirth, and J. Sinova, Phys. Status Solidi RRL {\bf 11}, 1700022 (2017).
\bibitem{wildes}A. R. Wildesdag, B. Roessli, B. Lebech, and K. W. Godfrey, J. Phys.: Condens. Matter {\bf 10}, 6417 (1998). 
\bibitem{kurosawa}K. Kurosawa, S. Saito, and Y. Yamaguchi, J. Phys. Soc. Jpn. {\bf 52}, 3919 (1983).
\bibitem{brec}R. Brec, Solid State Ionics {\bf 22}, 3 (1986).
\bibitem{saitoh}E. Saitoh, M. Ueda, H. Miyajima, and G. Tatara, Appl. Phys. Lett. {\bf 88}, 182509 (2006).
\bibitem{date}K. Okuda, {\it et al.} J. Phys. Soc. Jpn. {\bf 55}, 4456 (1986).
\bibitem{SM}See Supplemental Material at xxxx for further discussion and details.
%\bibitem{nakajima-jpsj}T. Nakajima {\it et al.} J. Phys. Soc. Jpn. {\bf 80}, 014714 (2011).
%\bibitem{ramazanoglu}M. Ramazanoglu {\it et al.} Phys. Rev. Lett. {\bf 107}, 067203 (2011).
%\bibitem{nakajima-prl}T. Nakajima {\it et al.} Phys. Rev. Lett. {\bf 114}, 067201 (2015).
%\bibitem{castro}A. C. Garcia-Castro, A. H. Romero, and E. Bousquet, Phys. Rev. Lett. {\bf 116}, 117202 (2016). 
\bibitem{otani}Y. Shiomi, {\it et al.} Appl. Phys. Lett. {\bf 104}, 242406 (2014).
\bibitem{Pt-seebeck}J. P. Moore and R. S. Graves, J. Appl. Phys. {\bf 44}, 1174 (1973).
\bibitem{barnard}R. D. Barnard, {\it Thermoelectricity in Metals and Alloys}, Taylor \& Francis Ltd. (1972) 
\bibitem{ferrotroidic}E. Ressouche, {\it et al.}, Phys. Rev. B {\bf 82}, 100408 (2010).
\bibitem{seki}S. Seki, {\it et al.} Phys. Rev. B {\bf 75}, 100403(R) (2007).
\end{thebibliography}
\end{document}